\documentclass[iop]{emulateapj}

\usepackage{epstopdf}

\newcommand{\kms}{\hbox{km s$^{-1}$}}
\newcommand{\msun}{\hbox{$M_\odot$}}

\newcommand{\atl}{ATLAS$^{\rm 3D}$}
\newcommand{\re}{\hbox{$R_{\rm e}$}}

\begin{document}

\title{Salpeter normalization of the Stellar Initial Mass Function\\ for massive galaxies at $z\sim1$}
\author{Shravan Shetty$^1$ and Michele Cappellari}
\affil{Sub-Department of Astrophysics, Department of Physics, University of Oxford, Keble Road, Oxford, OX1 3RH, UK \\
$^1$ Email: shravan.shetty@astro.ox.ac.uk}

\begin{abstract}

The stellar initial mass function (IMF) is a key parameter to study galaxy evolution.
Here we measure the IMF mass normalization for a sample of 68 field galaxies in the redshift range 0.7 to 0.9 within the Extended Groth Strip. To do this we derive total (stellar + dark matter) mass-to-light [$(M/L)_{\rm dyn}$] using axisymmetric dynamical models. Within the region where we have kinematics (about one half-light radius), the models assume: (i) that mass-follows-light, implying negligible differences between the stellar and total density profiles; (ii) constant velocity anisotropy ($\beta_{\rm z}\equiv1-\sigma_z^2/\sigma_R^2=0.2$); (iii) that galaxies are seen at the average inclination for random orientations (i.e. $i=60^\circ$, where $i=90^\circ$ represents edge-on). The dynamical models are based on anisotropic Jeans equations,  constrained by HST/ACS imaging and the central velocity dispersion of the galaxies, extracted from good-quality spectra taken by the DEEP2 survey. The population $(M/L)_{\rm pop}$ are derived from full-spectrum fitting of the same spectra with a grid of simple stellar population models. Recent dynamical modelling results by the \atl\ project and numerical simulations of galaxy evolution indicate that the dark matter fraction within the central regions of our galaxies should be small. This suggest that our derived total $(M/L)_{\rm dyn}$ should closely approximate the stellar $M/L$.
Our comparison of $(M/L)_{\rm dyn}$ and $(M/L)_{\rm pop}$ then imply that for galaxies with stellar mass $M_\ast\ga10^{11}$ \msun, the {\em average} normalization of the IMF is consistent with a Salpeter slope, with a substantial scatter. This is similar to what is found within a similar mass range for nearby galaxies.

\end{abstract}


\section{Introduction}

Knowledge of the stellar mass of galaxies has long been pivotal to the study and testing of galaxy formation and evolution theories, but an economical method for it's estimation has remained elusive to astronomers for decades \citep[see][for a review]{Courteau2014}. The main reason for this is the fact that the galaxy photometry doesn't sufficiently constrain the stellar mass, which is dependent on both the stellar populations of the galaxy and the stellar Initial Mass Function (IMF) of those populations. The IMF represents the distribution of stellar masses at a single star formation event. The form of the IMF is critical for estimating the mass of the galaxies and in understanding the stellar feedback and the chemical enrichment processes taking place within them. 

It is currently accepted, from direct star counts, that the IMF shape in our Milky Way can be approximated by a power-law $d\,N/d\,m \propto m^x$ with the \citet{salpeter1955} slope $x\approx-2.35$, above a stellar mass $m\ga0.5$ \msun\ and a more shallow slope below $m \la 0.5\msun$ \citep{kroupa2001,chabrier2003}. Though this IMF seems universal within the Milky Way its applicability to external galaxies has been uncertain for decades (see \citealt{Bastian2010} and \citealt{kroupa2013} for recent reviews).

Early attempts, comparing the dynamical mass-to-light ratio M/L to the value inferred from stellar population synthesis, concluded that at least some spiral galaxies \citep{bell2001,kassin2006,bershady2011,Dutton2011,brewer2012} and ETGs \citep{cappellari2006,Ferreras2008,Thomas2011} require an overall Kroupa/Chabrier mass normalization of the IMF like the Milky Way. By studying the variation of the average M/L of galaxies in clusters at different redshifts, \cite{renzini2006} concluded that a Salpeter IMF slope was required for the mass range of $1<M<1.4$ \msun.

More recently \citet{vandokkum2010,vanDokkum2011} studied near-infrared spectral features, where the contribution from dwarf stars is easier to measure. They concluded that massive elliptical galaxies have a dwarfs-dominated IMF, unlike the Milky Way. 
\citet{auger2010b} studied the IMF by comparing stellar masses from strong gravitational lensing to stellar masses from population synthesis. Their models assume massive ETGs are spherical, satisfy a power law dependence of IMF normalization with galaxy mass, have dark halos following the mass-concentration relation predicted by simulation and the total to stellar mass derived by halo abundance matching. They concluded massive ETGs as a class have a Salpeter-like mass normalization of the IMF.

\citet{cappellari2012nature} used more general axisymmetric dynamical models allowing for the IMF, the stellar profiles, the halo slope and the stellar fraction to freely vary in different galaxies to reproduce the data. The models were constrained by integral-field stellar kinematics for a sample of 260 ETGs spanning a large range of masses and $\sigma$. They found a systematic trend in the IMF normalization. The average IMF varies from Kroupa/Chabrier to Salpeter and heavier with increasing $\sigma$, which was shown to trace the bulge fraction \citep{atlas3d20}.

A number of recent works appear so far consistent with a systematic IMF variation. These used either spectral synthesis to study absorption features \citep[e.g.][]{Spiniello2012,Conroy2012,Smith2012,Ferreras2013,LaBarbera2013}, dynamical scaling relations \citep{Dutton2013} or approximate spherical dynamical models \citep{tortora2013}. However quantitative agreement between the different approaches has not yet been achieved and full consensus has not been reached \citep[e.g.][]{Maraston2013}.

The study of IMF in galaxies at high redshift is so far dominated by dynamical mass derivations using the virial estimator \citep[e.g.][]{renzini2006,vandeSande2013}. In this study, we constrain the IMF normalization at redshift $z\sim1$ using axisymmetric dynamical models and compare the results with findings in the local universe. This is one of the few studies to attempt dynamical modelling at higher redshift \citep{vanderMarel2007,cappellari2009} but is the first to place constraints on the IMF normalization of galaxies at high redshift. 

In section 2, we give a description of the data used for this study and a description of the galaxy set. We describe our methodology in section 3, while we present our results in section 4. In section 5, we discuss the results. We used the following cosmological constants: $\Omega_{\rm m}=0.3$, $\Omega_{\rm \Lambda}=0.7$, $H_{\rm 0}=70$ \kms\ Mpc$^{-1}$.

\section{Data and Sample}

\subsection{Spectral Data}

The 1-D spectrum of the galaxies was obtained from the DEEP2 spectrographic survey \citep{deep2}. It is a magnitude limited, $R_{\rm AB} \leq$ 24.1, galaxy redshift survey. In this study, we use the Extended Groth Strip \citep{groth1994} field of the survey due to the availability of HST imaging \citep{aegis}. The data was taken by the DEIMOS multi-object spectrograph, mounted on the Keck-2 telescope, with a observed wavelength range of 6500-9100\AA, in a spectral resolution of $R\sim6000$ at 7800\AA. The typical total exposure time for each galaxy is 1 hour, with average seeing of $0\farcs85$. The 2-D spectra of the galaxies were observed using slitlets of dimensions 1$''\times7''$. 1-D spectra of the galaxies were then derived using the Boxcar technique, within extraction windows of $\sim1\arcsec\times1\arcsec$. Due to the design of the survey, the spectra of the objects are divided into 2 halves. For our study, we use the blue half of the wavelength range so as to observe the CaK and G stellar lines for our given redshift range.

\subsection{Photometric Data}

A HST/ACS survey of the EGS was done by the GO program 10134 [P.I: M. Davis] \citep{aegis}. The survey was done with the F606W (V) and F814W (I) filters, at 5 $\sigma$ magnitude limits of $V_{\rm F606W}=26.23(AB)$ and $I_{\rm F814W}=25.61(AB)$. The observations were done in a four point dither pattern, which was processed by the STSDAS MultiDrizzle package to produce a final mosaic of the field with a pixel scale of $0\farcs03$.

\subsection{Selection}

The DEEP2 survey has observed $\sim49,000$ galaxy spectra. For this study, our initial selection criteria are: the galaxies have reliable redshift between $0.7<z<0.9$ and have high-resolution HST photometry. This meant selecting galaxies with  ``secure'' and ``very secure'' redshifts in the DEEP2 catalog. These criteria reduced the sample to $\sim1,350$. 

The galaxy spectra were then logarithmically rebinned to 60 \kms\ per spectral pixel and a signal-to-noise $(S/N)>3$ cut-off was applied. We visually inspected the spectra after their initial spectral fits (Section \ref{sec:ppxf}) and retained galaxies showing clear and well-fitted stellar absorption lines. This reduced the sample to $\sim200$. 

After the photometric analysis (Section \ref{sec:jam}), a visual inspection of the contour fits of the galaxy photometry and the model was done to ensure that the models accurately reproduce the galaxy photometry, removing galaxies with non-axisymmetric features, e.g. disturbed morphology or strong dust. This reduced our sample to 87. 

Another selection criterion implemented on our galaxy sample was that of luminosity-weighted age of the galaxies, derived by identifying the best-fitting single stellar population model with solar metallicity via full-spectrum fitting. Galaxies with best-fitting template age \textless1.2 Gyr were removed. A similar selection was made by \citet{cappellari2012nature,atlas3d20} who found that young galaxies have strong stellar population gradients, hence breaking the assumption of constant stellar M/L. Finally we removed galaxies with strong evidence for multiple population using full-spectrum fitting (Section \ref{sec:pop}), leading to a final sample of 68 galaxies.

These galaxies consists of massive ETGs (stellar masses $10^{11}\la M_\ast/\msun \la 10^{12}$), except for seven galaxies at lower masses. More information on these galaxies will be given in a follow up paper.

\begin{figure*}
	\centering
	\plotone{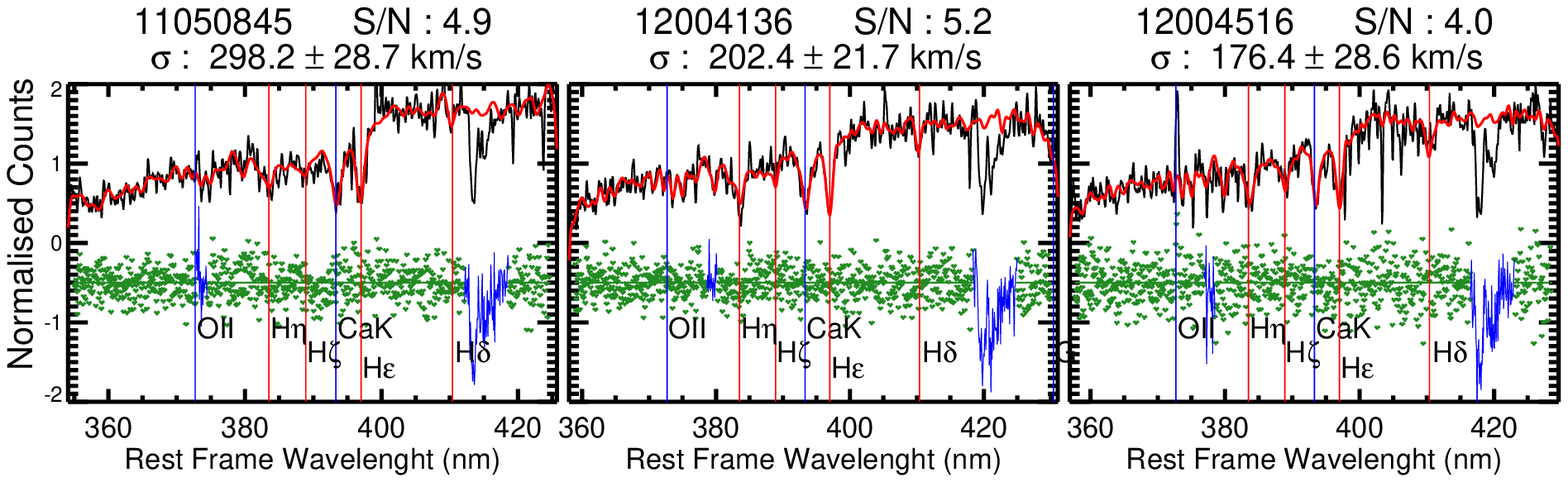}
	\plotone{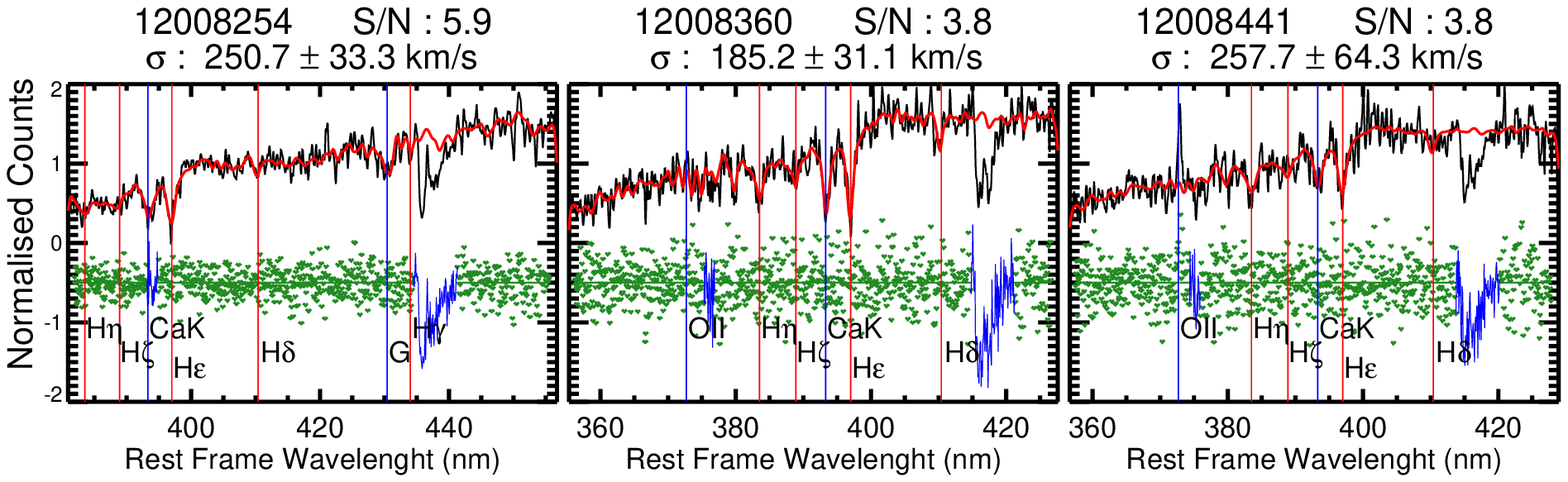}
	\plotone{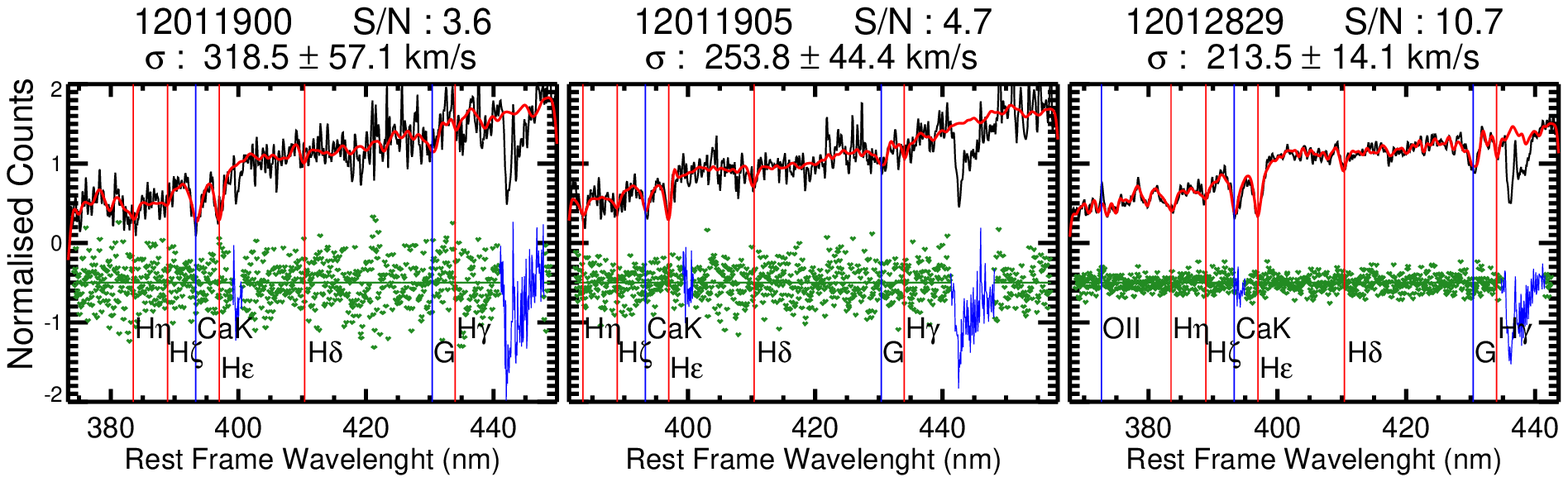}
	\caption{Summary plots of spectrum fitting for 9 galaxies. The galaxy spectrum is shown in black, while the best-fit is in red. The galaxy spectrum has been rebinned to 60 \kms spectral pixel resolution and gauss-smoothed to make the absorption features clearer to the eye. The green dots below the galaxy spectrum are the residuals of the fit, offset by -0.5 in the interest of clarity, while the blue features are the masked pixels. The portion of the spectrum around $\sim3800\AA$ and $\sim4100\AA$ are the telluric features that have been masked.}
	\label{Spec}
\end{figure*}

\begin{figure*}
	\centering
	\plotone{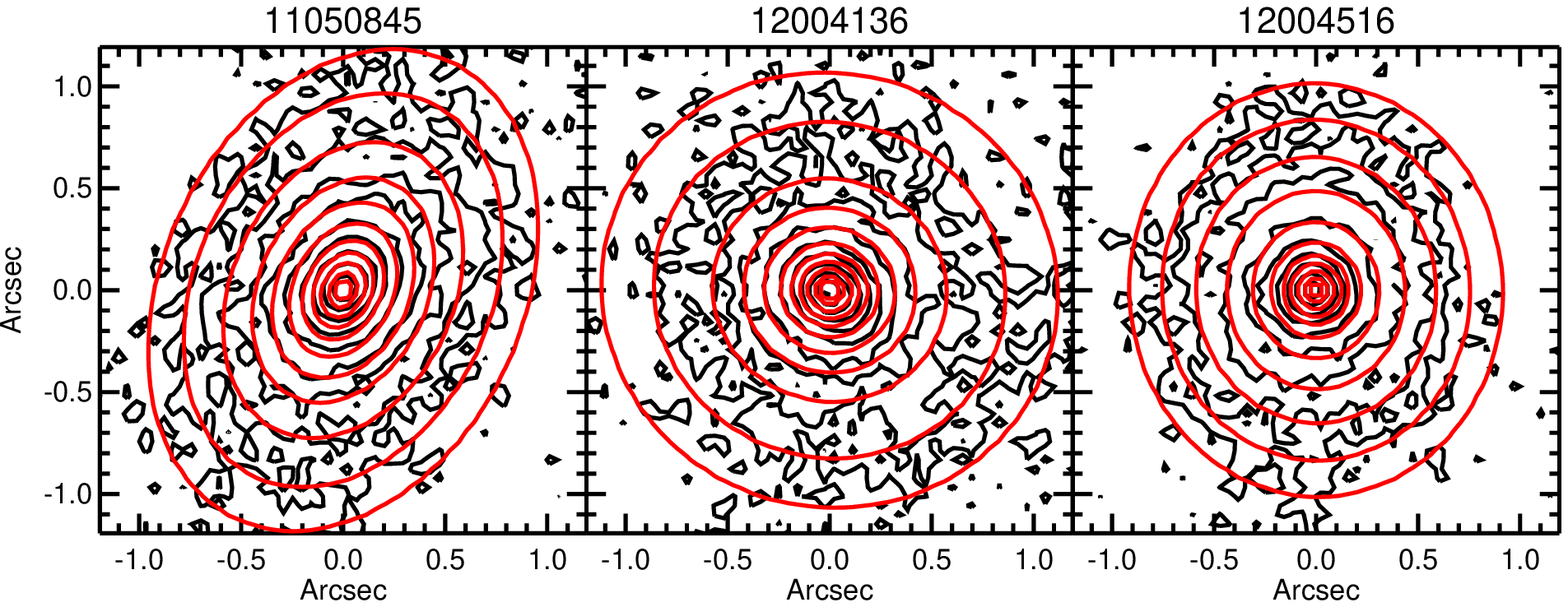}	
	\plotone{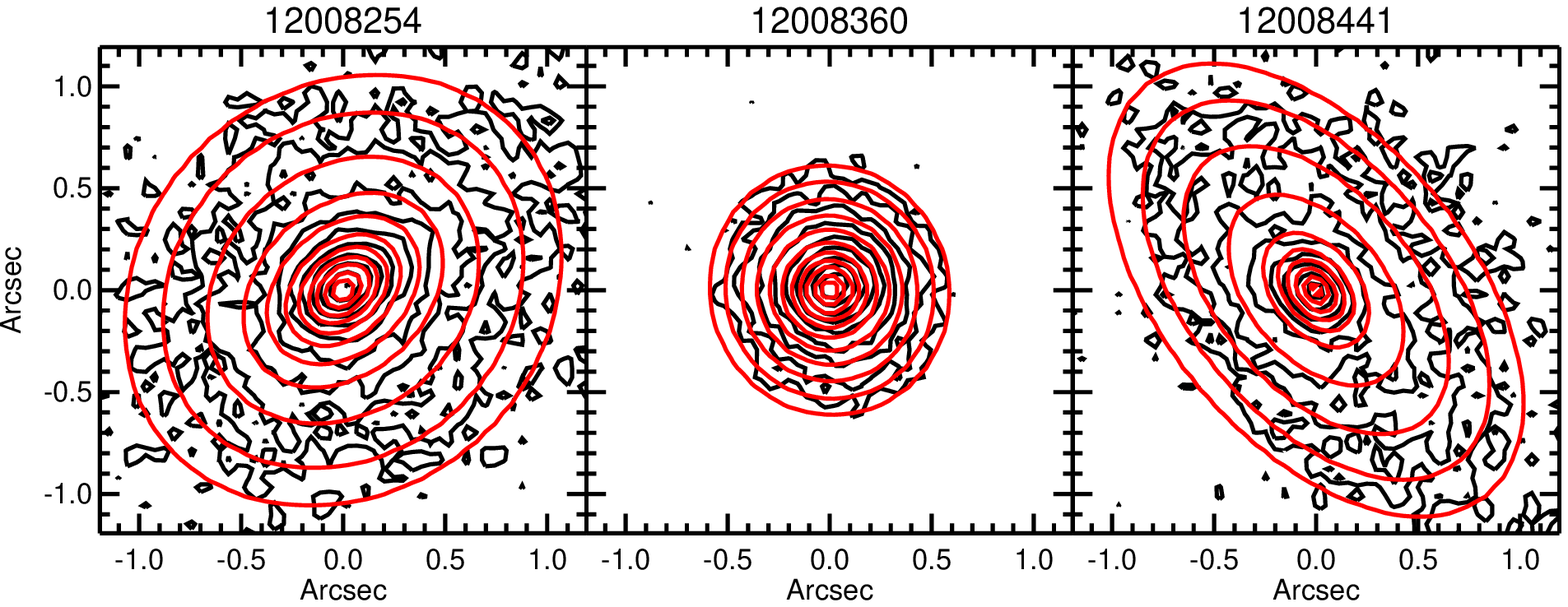}
	\plotone{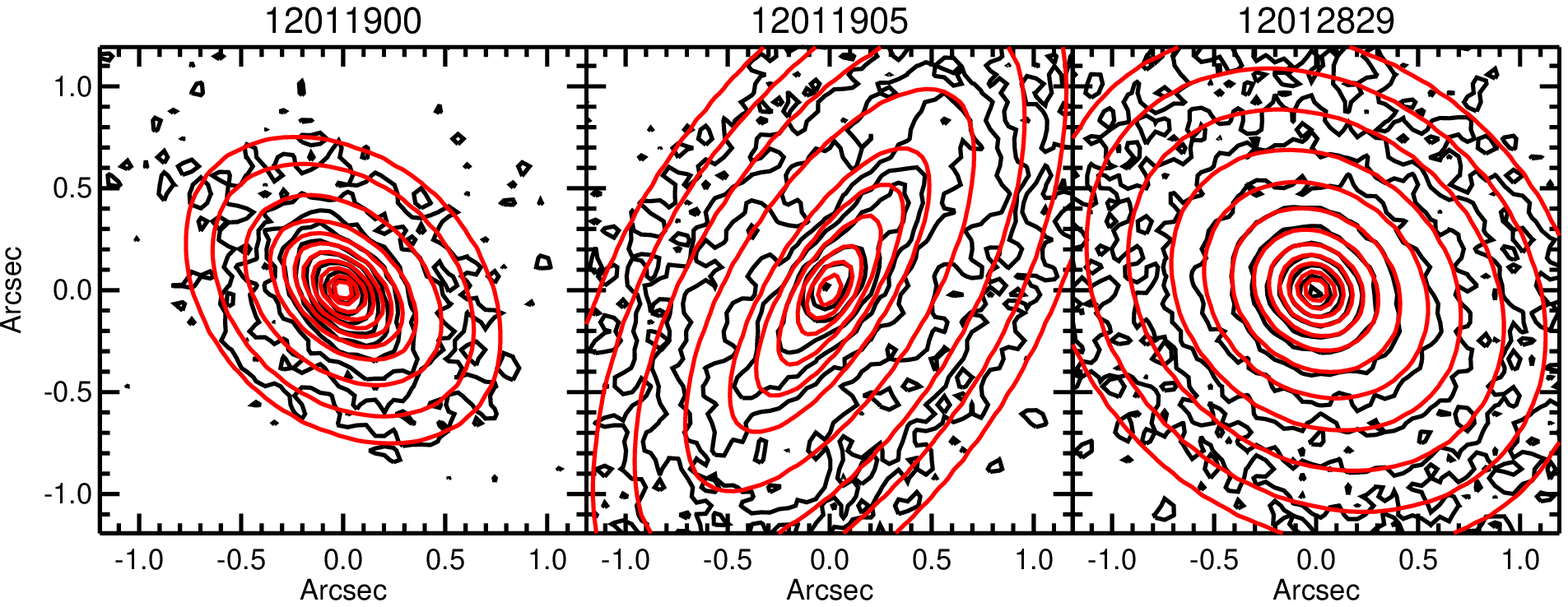}
	\caption{Summary plots of the MGE fits for 9 galaxies. The black contour lines represent the observed light distribution of the galaxies, while the red represents the model. Since we calculate the $(M/L)_{\rm JAM}$ for the inner regions of the galaxies, it is important to reproduce the inner light profile accurately.}
	\label{Contour}
\end{figure*}

\section{Methods}

\subsection{Velocity dispersion measurements}
\label{sec:ppxf}

To create dynamical models for our galaxies, we first extracted reliable velocity dispersion for the galaxies, using the pPXF code\footnote{Available at: http://purl.org/cappellari/software} \citep{ppxf}. The spectrum fitting was done with a subset of 53 stellar spectra taken from the Indo-US Library of Coud\'{e} Feed Stellar Spectra Library \citep{indo-uslibrary}. These spectra were selected so that: (i) each spectrum is gap-free and (ii) the subset is a good representation of the library's atmospheric parameter range ($T_{\rm eff}$ vs $[Fe/H]$). During the fitting process, the telluric features in the galaxy spectra were masked, along with any significant gas emission feature. Examples of spectral fits are given in Fig.~\ref{Spec}.

A bootstrapping technique was used to estimate the real errors on the derived single aperture velocity dispersions. The technique involved re-sampling the residuals between the observed spectrum and an initial best fit. The bootstrapped spectra are then fitted with an ``Optimal'' template which is the summation of the templates used for the initial fit of the spectrum, multiplied with their respective weights. The process was iterated 500 times, and the distribution of the stellar kinematic values was used to estimate the velocity dispersion and its error. 

We compare our derived velocity dispersion with that of \citet{fernandezlorenzo2011}, for our 40 common galaxies, using the LTS\_LINEFIT routine\footnotemark[1], modified to constrain the slope to 1. We find that our velocity dispersion are consistent with those of \citet{fernandezlorenzo2011} with negligible offset of $0.013\pm0.013$ dex and an observed rms scatter of 0.094 dex.

\subsection{Calculating Dynamical M/L}
\label{sec:jam}

The photometry of our galaxies was parametrized using a Multi-Gaussian Expansion \citep{Emsellem1994}, using the MGE\_FIT\_SECTORS$^1$ code of \citet{cappellari2002}. To ensure that our MGE fits reflect the underlying mass distribution of galaxies, we adopt the technique used by \citet{Scott2013} (Section 3.2.1). We also force the surface brightness profile at the largest observed radius to have an outer slope of $R^{-4}$. Examples of MGE fits are shown in Fig.~\ref{Contour}.

We create our dynamical models using the JAM method$^1$ \citep{jam} which solves the anisotropic Jeans equations for a given light and mass profile of axisymmetric galaxies, while allowing for orbital anisotropy. The code also allows for rigorous application of seeing and aperture effects.

We adopt this method over the more commonly used virial estimates based on S\'{e}rsic spherical Jeans models since spherical models cannot reproduce the variety in the stellar kinematics of galaxies \cite[Fig. 1]{atlas3d15}. In addition, \citet[Fig.14]{atlas3d15}, show that virial estimates suffer from large scatter and potential biases in the absolute M/L normalization, which is critical for our work.

We assume for all galaxies the average inclination  $i=60^\circ$ for random orientations ($i=90^{\circ}$ being edge-on). For a few galaxies for which this inclination is not allowed by the deprojection of the MGE model, we assign the lowest allowed inclination. We adopt an anisotropy $\beta_{\rm z}\equiv1-\sigma_z^2/\sigma_R^2=0.2$ which is the typical value, within one half-light radius \re, for early type galaxies in the local universe \citep{sauron10,Gerhard2001}. Previous studies have shown the $M/L$ to be very weakly sensitive to the adopted inclination \citep[Fig.~4]{cappellari2006}. But our quoted errors do not include the effect of our constant anisotropy assumption.

Given our lack of spatially resolved data, we are unable to separate luminous and dark matter in our galaxies. Hence we use mass-follow-light models, which were shown to robustly recover the {\em total} $M/L$ within 1\re, even when dark matter is present \citep{atlas3d15}, when the data extend out to the same radius. The dynamical $M/L$ of our galaxies is given by $(M/L)_{\rm JAM}=(\sigma/V_{\rm rms})^2$ (Table~1), where $\sigma$ is measured within the DEEP2 1\arcsec\ slit (Table 1) and $V_{\rm rms}$ is the value predicted by the JAM model within the same aperture for $M/L=1$.

The extensive modelling study of \citet{atlas3d15} measure a low dark matter fraction  ($<30\%$), within $1R_{\rm e}$, for nearby galaxies of similar mass as ours. Numerical simulations by \citet{Hilz2013} provide a general argument as to why dark matter should further decrease with increasing redshift. These results indicate that the dark matter fraction in the central regions of our galaxies should be negligible, hence indicating that our $(M/L)_{\rm JAM}$  should closely approximate the stellar $M/L$. Without this assumption, our $M/L$ would only provide an upper limit to the IMF mass normalization.

We use the parametrization of the photometry and the redshift of the galaxies to derive their absolute magnitude. Since the F814W filter observes the rest frame B band for galaxies at redshift $\sim0.8$, all magnitudes and $(M/L)$s presented in this study are in the B (Vega) system. 

\begin{table*}
\begin{center}
\caption{Results of dynamical models and multiple population fitting. The table is published in its entirety in the electronic edition of the letter. A portion is shown here for guidance regarding its form and content.}
\begin{tabular}{ccccccccc}
\hline\hline
{DEEP2 ID No.} &
{Redshift} &
{$M_{\rm B}$} &
{$R_{\rm e}$} &
{$\sigma$} &
{$\Delta\sigma$} &
{$(M/L)_{\rm JAM}$} &
{$(M/L)_{\rm Sal}$} \\
{} &
{} &
{} &
{(\arcsec)} &
{(km/s)} &
{(km/s)} &
{(\msun/$L_{\odot B}$)} &
{(\msun/$L_{\odot B}$)} \\
{(1)} &
{(2)} &
{(3)} &
{(4)} &
{(5)} &
{(6)} &
{(7)} &
{(8)} \\ 
\hline
11050845 & 0.840 & -21.76 & 0.40 & 298 & 28 & 6.10 & 3.55 \\
12004136 & 0.812 & -21.62 & 0.50 & 202 & 21 & 3.09 & 2.92 \\
12004516 & 0.820 & -21.44 & 0.44 & 176 & 28 & 2.85 & 1.97 \\
12008254 & 0.745 & -21.51 & 0.41 & 250 & 33 & 4.92 & 4.32 \\
12008360 & 0.828 & -20.97 & 0.18 & 185 & 31 & 2.63 & 2.39 \\
12008441 & 0.832 & -21.34 & 0.53 & 257 & 64 & 9.26 & 1.79 \\
12011900 & 0.719 & -20.88 & 0.23 & 318 & 57 & 9.75 & 4.46 \\
\hline
\end{tabular}
\label{data}
\tablecomments{
Column (1): DEEP2 galaxy identifier.
Column (2): DEEP2 estimated redshift.
Column (3): Absolute B (Vega) band magnitude calculated using MGE.
Column (4): Effective radius as derived from MGE (without re-scaling).
Column (5): Aperture velocity dispersion.
Column (6): Error on aperture velocity dispersion.
Column (7): Dynamical B-band mass-to-light ratio with a median error of 0.11 dex.
Column (8): Population B-band mass-to-light ratio with a scatter of 0.08 dex.
}
\end{center}
\end{table*}

\subsection{Calculating Stellar Population M/L}
\label{sec:pop}

We derive the population M/L of the galaxies using the MILES stellar evolutionary models \citep{milesmodels}. These models use the MILES empirical stellar spectral library \citep{mileslibrary} to derive single age and metallicity models for the entire optical wavelength range. We use the Salpeter IMF as the reference IMF for this study. 

The full-spectrum fittings were done using pPXF with a template grid of 40 logarithmically spaced ages, between 0.089 to 7.9 Gyrs, for metallicities [M/H] of -0.4, 0.00 and 0.22. The upper limit of the age was constrained to the age of the universe at the adopted redshift while the metallicity range is justified by the results of \citet{schiavon2006} who find that galaxies at $0.7<z<0.9$ have nearly solar metallicities. The $(M/L)_{\rm Salpeter}$ in the B (Vega) band was calculated as
\begin{equation}
(M/L_{\rm B})_{\rm Sal}=\sum_{j} \frac{w_{j}M_{\star}^{\rm nogas}}{w_{j}L_{\rm B}},
\end{equation}
where $w_{j}$ is the weight of each template, $M_{\star}^{\rm nogas}$ is the mass of stars and stellar remnants, and $L_{\rm B}$ is the B (Vega) band luminosity of the model. 

To test the robustness of our results to model selection, the $(M/L)_{\rm Salpeter}$ was calculated using 2 different template sets: (i) unregularized fitting with solar metallicity templates and (ii) a regularized fitting with the entire SSP model grid. The M/L derived from these template sets are offset by only 0.07 dex and have an intrinsic scatter of 0.08 dex. We find that our results are robust to variations of this scale and so we present our results here using the latter set.

\begin{figure}
	\plotone{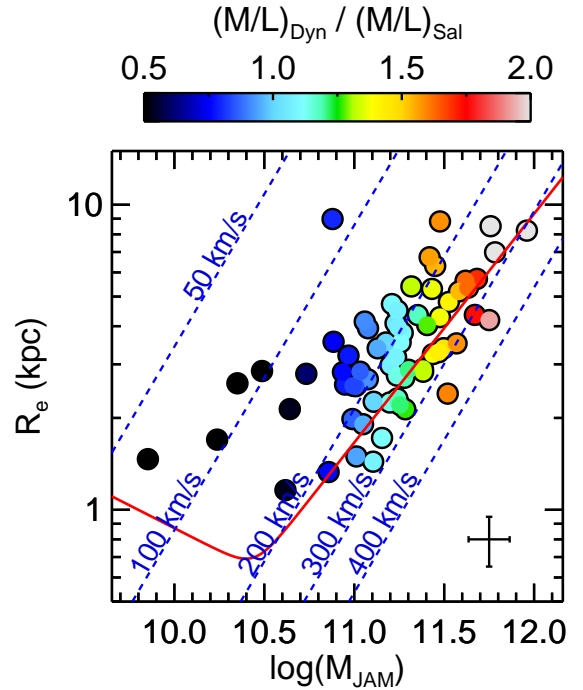}
	\caption{Mass Plane ($R_{\rm e}$ vs Stellar Mass) of our final galaxy set, with $R_{\rm e}$ rescaled using the offset observed between $(M/L)_{\rm JAM}$ and the virial estimate. The thick red line represents the Zone of Exclusion from \citet{atlas3d20}, which approximates the 99\% boundary for the nearby galaxy population. The colored diagonal dashed lines are predicted lines of constant velocity dispersion according to the scalar virial equation. Part of the trend in mass normalization with mass must be due to the correlation between $M_{\rm JAM}$ and $(M/L)_{\rm Dyn}$.}
	\label{MassPlane}
\end{figure}

\begin{figure}
	\plotone{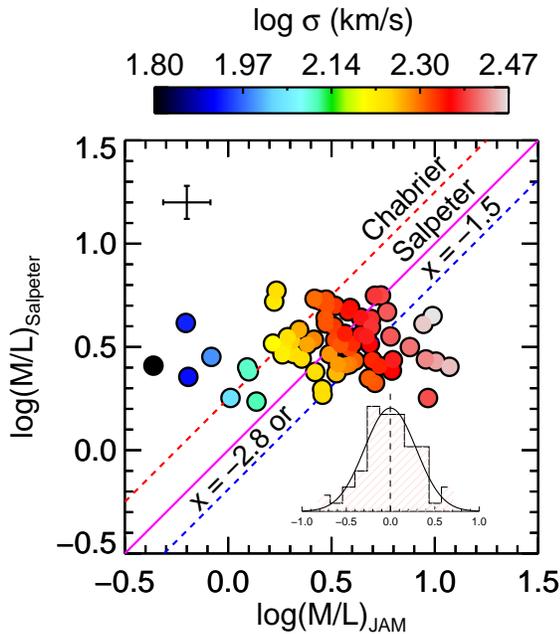}
	\caption{log$(M/L_{\rm B})_{\rm JAM}$ versus log$(M/L_{\rm B})_{\rm Sal}$. The Thick Magenta line represents the Salpeter IMF. The Dashed Red line represents the Chabrier IMF, while the Dashed Blue line represents a single power-law with a slope of -2.8. The points are color coded to the velocity dispersion smoothed with LOESS method of \citet{cleveland1979}.Though there is an indication of a trend of the mass normalization with the velocity dispersion of galaxies, part of this must be due to the correlation between $\sigma$ and $(M/L)_{\rm JAM}$. A representative error bar is shown at the top-left. The histogram on the bottom-right shows the log$((M/L)_{\rm Dyn}/(M/L)_{\rm Sal})$. The best-fitting Gaussian peaks at 0.00 and has a dispersion of 0.29. We note that part of the trend of velocity dispersion with $(M/L)_{\rm JAM}$ is likely due to correlated errors.}
	\label{atlas3d20}
\end{figure}

\section{Result}
 
In this study, we derive accurate total dynamical $(M/L)_{\rm JAM}$ and stellar population $(M/L)_{\rm Salpeter}$ for 68 galaxies. In Table~\ref{data}, the reader will find the list of physical parameters derived in the study.

Our results allow us to compare the $M/L$ from detailed models with the virial estimate: 
\begin{equation}
(M/L)_{\rm virial}= 5.0\times \frac{R_{\rm e}\sigma^{2}_{e}}{G\, L},
\label{virialeq}
\end{equation}
where $\sigma_{\rm e}$ is the measured velocity dispersion within \re\ \citep{cappellari2006}. In our study, we have measured the \re\ using MGE and use the aperture correction for the velocity dispersion as given in of \citet[Eq.~(1)]{cappellari2006}. We use a virial coefficient of 5.0 since \citet[Fig. 13]{atlas3d15} shows that it is applicable to virial mass estimations that use MGE derived \re. We find that the equation~(\ref{virialeq}) under estimates the dynamical masses by 0.16 dex with an rms scatter of 0.04 dex.

We investigated the origin of the offset via a process of elimination. Since the virial estimate, using the MGE derived sizes better reproduces the dynamical mass of galaxies than a non-homology dependent virial coefficient \citep[Fig. 14]{atlas3d15}, we argue that the structural evolution of galaxies can not account for this offset. An under-estimation in velocity dispersion and luminosity will have the same effect on both quantities, apart from effects of seeing, which is only accounted for in the models. Hence our investigation suggests that the offset observed is mostly caused by an under-estimation of the \re, specifically due to shallow observations of galaxy photometry and our non-extrapolated radii.

In Fig.~\ref{MassPlane}, we show the mass plane ($R_{\rm e}$ vs Stellar Mass) of the galaxies. The \re\ used in the plot is rescaled to match the $(M/L)_{\rm JAM}$ and  $(M/L)_{\rm virial}$ estimates. It is an attempt to recover the true \re\ of the galaxies, but this is assumption dependent. Regardless, none of our results on the IMF mass normalization are dependent on this. A number of galaxies appear slightly smaller than local galaxies, and lie below the zone-of-exclusion of Fig.~1 of \citet{atlas3d20}. This illustrates that our $S/N$ selection criterion tend to select dense and massive galaxies.

By comparing the M/L calculated via the two independent methods, dynamical modeling and stellar populations, we are able to study the IMF normalization of the galaxies. In Fig.~\ref{atlas3d20} we have plotted the M/L calculated by the two methods, along with a color code to represent velocity dispersion. The velocity dispersion is smoothed using the LOESS method of \citet{cleveland1979} \footnote{As implemented in the CAP\_LOESS\_2D routine of \citet{atlas3d15}, available from http://purl.org/cappellari/software} to ease comparison of the plot with the results of the local universe, as done in \citet{atlas3d20}.

The analysis of our results show that, on {\em average}, massive ($10^{11}\la M_\ast/\msun\la10^{12}$) galaxies at $z\sim0.8$ are more consistent with a Salpeter-like, than a Milky Way-like, IMF mass normalization provided our assumption of negligible dark matter in the central regions of the galaxies is valid. The IMF scatter appear larger than our error estimate, suggesting it must be intrinsic. However the extreme cases must be driven by systematics in our $\sigma$ determinations. Fig.~\ref{atlas3d20} can be directly compared to Fig. 11 of \citet{atlas3d20} representing the local galaxy sample. The two plots look qualitatively similar, and in both cases they are characterized by a nearly complete lack of correlation between the dynamical and population M/L for galaxies with the largest $\sigma$. 

\section{Summary}

We used the JAM modeling method to derive dynamical $(M/L)_{\rm JAM}$ for 68 massive galaxies ($M_\ast=10^{11}-10^{12}$ \msun) at redshifts $z=0.7-0.9$. Our results show that using the virial equation at high redshifts can lead to severe under-estimation in the dynamical masses of galaxies, which is likely due to the under-estimation of the $R_{\rm e}$. The under-estimation must depend on the depth of the photometry as well as on the adopted extrapolation of the profiles, making the virial estimates assumption dependent \citep[see also][]{atlas3d15}. The $(M/L)_{\rm JAM}$ doesn't suffer for this problem as it only requires the knowledge of the light and mass distribution within a region comparable to where the stellar kinematics are available.

Our result suggests that stellar population estimates of stellar masses of massive galaxies at high redshift should assume a Salpeter IMF normalization for more accurate results, instead of the Kroupa/Chabrier IMFs which are often adopted. This IMF normalization is the same inferred for nearby galaxies of similar masses.

\bibliographystyle{apj}


\end{document}